\documentclass[
prl,
nofootinbib,
twocolumn,
superscriptaddress,
showpacs
]{revtex4}

\usepackage{graphicx}
\usepackage{epsfig}
\usepackage{latexsym}
\usepackage{amsmath}
\usepackage{amsfonts}
\usepackage{amsxtra}

\newcommand{\msbar}{\overline{\mbox{{\sc ms}}}}

\newcommand{\lwrsim}{\raise0.3ex\hbox{$<$\kern-0.75em\raise-1.1ex\hbox{$\sim$}}}
\def\krto{ {\,\,\lower .8ex\hbox {$\longrightarrow \atop k \rightarrow 0$}\,\,}}

\def\bea{\begin{eqnarray} }
\def\beq{\begin{eqnarray} }

\def\eea{\end{eqnarray}}
\def\eeq{\end{eqnarray}}

\def\eq#1{Eq.~(\ref{#1})}

\newcommand{\ams}{\alpha_{\overline{\rm MS}}}
\newcommand{\Lms}{\Lambda_{\overline{\rm MS}}}

\begin{document} 

\title{The strong running coupling at $\tau$ and $Z_0$ mass scales from lattice QCD}

\author{(ETM Collaboration) B.~Blossier}
\author{Ph.~Boucaud} 
\affiliation{Laboratoire de Physique Th\'eorique, 
Universit\'e de Paris XI; B\^atiment 210, 91405 Orsay Cedex; France}
\author{M.~Brinet}
\affiliation{LPSC, CNRS/IN2P3/UJF; 
53, avenue des Martyrs, 38026 Grenoble, France}
\author{F.~De Soto}
\affiliation{Dpto. Sistemas F\'isicos, Qu\'imicos y Naturales, 
Univ. Pablo de Olavide, 41013 Sevilla, Spain}
\author{X.~Du}
\affiliation{LPSC, CNRS/IN2P3/UJF; 
53, avenue des Martyrs, 38026 Grenoble, France}
\author{V.~Morenas}
\affiliation{Laboratoire de Physique Corpusculaire, Universit\'e Blaise Pascal, CNRS/IN2P3 
63177 Aubi\`ere Cedex, France}
\author{O.~P\`ene}
\affiliation{Laboratoire de Physique Th\'eorique, 
Universit\'e de Paris XI; B\^atiment 210, 91405 Orsay Cedex; France}
\author{K.~Petrov}
\affiliation{CEA Saclay, Irfu/SPhN; Orme des Merisiers, Bat 703; 91191 Gif/Yvette Cedex; France}
\author{J.~Rodr\'{\i}guez-Quintero}
\affiliation{Dpto. F\'isica Aplicada, Fac. Ciencias Experimentales; 
Universidad de Huelva, 21071 Huelva; Spain.}

\begin{abstract}

This letter reports on the first computation, from data obtained in lattice QCD with $u,d,s$ and $c$ quarks 
in the sea, of the running strong coupling via the ghost-gluon coupling renormalized 
in the MOM Taylor scheme. We provide with estimates of $\ams(m_\tau^2)$ and $\ams(m_Z^2)$ 
in very good agreement with experimental results.
Including a dynamical c quark makes safer the needed running of $\ams$.

\end{abstract}


\maketitle





\section{Introduction}

The confrontation of QCD, the theory for the strong interactions, with 
experiments requires a few inputs: one mass parameter for each quark species and 
an energy scale surviving in the limit of massless quarks, $\Lambda_{\rm QCD}$. 
This energy scale is typically used as the boundary condition to integrate 
the Renormalization Group equation for the strong coupling constant, $\alpha_S$. 
The value of the renormalized strong coupling at any scale, 
or equivalently $\Lambda_{\rm QCD}$, has to be fitted to allow the QCD phenomenology to account 
successfully for experiments. A description of many precision measurements of 
$\alpha_S$ from different processes and at different energy scales can be found in ref.~\cite{Bethke:2011tr}.  
The running QCD coupling can be alternatively obtained from lattice computations,  
where the lattice spacing replaces $\Lambda_{\rm QCD}$ as a 
dimensionful parameter to be adjusted from experimental inputs. This means that 
a lattice-regularized QCD can be a tool to convert 
the physical observation used for the lattice spacing calibration, 
as for instance a mass or a decay constant, into $\Lambda_{\rm QCD}$. A review of 
most of the procedures recently implemented to determine the strong coupling 
from the lattice can be found in ref.~\cite{Aoki:2010yq}.  We also quoted in 
ref.~\cite{Blossier:2011tf} many of the different methods proposed in 
the last few years.

The present ``{\it world average}'' for the strong coupling determinations~\cite{Nakamura:2010zzi}, 
usually referred at the $Z^0$-mass scale,  
is dominated by the lattice determination included in the average~\cite{Davies:2008sw}, 
as discussed in ref.~\cite{Bethke:2011tr}.  
Because of the importance of a precise and proper knowledge of the strong coupling for 
the LHC cross sections studies and its exploration of new physics, independent alternative lattice 
determinations are strongly required. The latter is specially true when different lattice actions 
and procedures are applied, to gain thus the best possible control on any source of systematic 
uncertainty. Furthermore, the current lattice results have been obtained by means of simulations 
including only two degenerate up and down sea quarks (${\rm N}_f$=2) or, as in ref.~\cite{Davies:2008sw}, 
also including one more ``tuned'' to the strange quark (${\rm N}_f$=2+1). Now, 
the European Twisted Mass (ETM) collaboration has started a wide-ranging program of 
lattice QCD calculation with two light degenerate twisted-mass flavours~\cite{Frezzotti:2000nk,Frezzotti:2003xj} 
and a heavy doublet for the strange and charm dynamical quarks (${\rm N}_f$=2+1+1)~\cite{Baron:2010bv,Baron:2011sf}. 
Within this ETM program, 
we have applied the method to study the running of the strong coupling, and so evaluate $\Lambda_{\rm QCD}$, 
grounded on the lattice determination of the ghost-gluon coupling in the so-called MOM Taylor 
renormalization scheme~\cite{Boucaud:2008gn,Blossier:2010ky}. We are publishing the results of 
this study in two papers: a methodological one~\cite{Blossier:2011tf}, where the procedure is  
described in detail along with some results, and this short letter aimed to update and emphasize 
the phenomenologically relevant results. In particular, as far as the lattice gauge fields 
with 2+1+1 dynamical flavours which we are exploiting provide with a very realistic simulation 
of QCD at the energy scales for the $\tau$ physics, we are presenting here the estimate for the 
coupling at the $\tau$-mass scale and directly comparing with the one obtained 
from $\tau$ decays. It should be noted that including the dynamical charm quark makes also safer 
the running up to the $Z^0$-mass scale.


\section{The strong coupling in Taylor scheme}
\label{sec:proc}

The starting point for the analysis of this letter shall be the Landau-gauge running strong coupling 
renormalized in the MOM-like Taylor scheme,  
\beq\label{alpha} 
\alpha_T(\mu^2) \equiv \frac{g^2_T(\mu^2)}{4 \pi} = \lim_{\Lambda \to \infty} 
\frac{g_0^2(\Lambda^2)}{4 \pi} G(\mu^2,\Lambda^2) F^{2}(\mu^2,\Lambda^2) \ ,
\eeq
obtained from lattice QCD simulations. $F$ and $G$ stand for the 
form factors of the two-point ghost and gluon Green functions (dressing functions). 
The procedure to compute the coupling defined by \eqref{alpha}, and from it to perform an estimate of
$\Lambda_{\msbar}$, is described in very detail in refs.~\cite{Boucaud:2008gn,Blossier:2010ky}. 
We recently applied this in ref.~\cite{Blossier:2011tf} to compute $\Lambda_{\msbar}$ 
from ${\rm N}_f$=2+1+1 gauge configurations for several bare couplings ($\beta$), 
light twisted masses ($a\mu_l$) and volumes. The prescriptions applied for the appropriate elimination 
of discretization artefacts, as the so-called $H(4)$-extrapolation 
procedure~\cite{Becirevic:1999uc}, 
were also carefully explained in ref.~\cite{Blossier:2011tf}. 
After this, we are left with the lattice estimates of the 
Taylor coupling, computed over a large range of momenta, that 
can be described above around 4 GeV (see Fig.~\ref{fig:plotalphaT}) by  
the following OPE formula\cite{Blossier:2010ky}: 
\beq\label{alphahNP}
\alpha_T(\mu^2)
&=& 
\alpha^{\rm pert}_T(\mu^2)
\ 
\left( \rule[0cm]{0cm}{0.85cm}
 1 + \frac{9}{\mu^2} \
R\left(\alpha^{\rm pert}_T(\mu^2),\alpha^{\rm pert}_T(q_0^2) \right) \right.
\nonumber \\
&\times& \left. \left( \frac{\alpha^{\rm pert}_T(\mu^2)}{\alpha^{\rm pert}_T(q_0^2)}
\right)^{1-\gamma_0^{A^2}/\beta_0} 
\frac{g^2_T(q_0^2) \langle A^2 \rangle_{R,q_0^2}} {4 (N_C^2-1)}
\right) , 
\eeq
where  $1-\gamma_0^{A^2}/\beta_0 = 27/100$ for $N_f=4$~\cite{Gracey:2002yt,Chetyrkin:2009kh}. 
$R(\alpha,\alpha_0)$ for $q_0=10$ GeV (see Eq.(6) of \cite{Blossier:2011tf}) is 
obtained as explained in the appendix of ref.~\cite{Blossier:2010ky}
%
. 
The purely perturbative running in \eq{alphahNP} 
is given up to four-loops by integration of the $\beta$-function~\cite{Nakamura:2010zzi},  
where its coefficients are taken to be defined 
in Taylor-scheme~\cite{Chetyrkin:2000dq,Boucaud:2008gn}. Thus, $\alpha_T^{\rm pert}$ 
depends only on $\ln{(\mu^2}/\Lambda_T^2)$. 
This however allows to fit both $g^2 \langle A^2 \rangle$ and $\Lambda_T$, 
the $\Lambda_{\rm QCD}$ parameter in Taylor scheme, through the comparison of the prediction given by 
\eq{alphahNP} and the lattice estimate of Taylor coupling. The best-fit of \eq{alphahNP} 
to the lattice data published in ref.~\cite{Blossier:2011tf} provided with the 
estimates that can be read in Tab.~\ref{tab:bestfit}.
In this letter, we complete the previous analysis by including an ``{\it ad-hoc}'' correction to account 
for higher power corrections (see Fig.~\ref{fig:plotalphaTPhi}) that allows to extend the fitting window down to $p \simeq 1.7$ GeV  
and also apply the so-called {\it plateau} method to determine the best-fit~\cite{Boucaud:2008gn}.  
Furthermore, in addition to the lattice ensembles of gauge configurations described in ref.~\cite{Blossier:2011tf},
we study 60 more at $\beta=2.1$ ($a\mu_l=0.002$) and three new ensembles of 50 configurations at 
$\beta=1.9$ and $a\mu_l=0.003,0.004,0.005$ to peform a chiral extrapolation for the ratios of lattice spacings.  
We get: $a(2.1,0.002)/a(1.9,0)=0.685(21)$. 
The lattice scale at $\beta=1.9,1.95,2.1$ is fixed by ETMC 
through chiral fits to lattice pseudoscalar masses and decay constants, 
where 270 $\lwrsim \; m_{\rm PS} \; \lwrsim$ 510 MeV, that are required 
to take the experimental $f_\pi$ and $m_\pi$ at the physical point~\cite{Baron:2010bv,Baron:2011sf}; 
{\it e.g.}~: $a(1.9,0)=0.08612(42)$ fm.

\section{The Wilson OPE coefficient and the higher-power  corrections}

The OPE prediction for $\alpha_T$ given by \eq{alphahNP} is dominated by 
the first correction introduced by the non-vanishing dimension-two Landau-gauge gluon 
condensate~\cite{Boucaud:2001st,
Boucaud:2001qz,Boucaud:2005xn,Gubarev:2000nz,Dudal:2005na,RuizArriola:2004en}, 
where the Wilson coefficient is applied at the ${\cal O}(\alpha^4)$-order. 
In the previous methodological paper~\cite{Blossier:2011tf}, we provided with 
a strong indication that the OPE analysis is indeed in order: it was clearly 
shown that the lattice data could be only explained by including non-perturbative contributions 
and that the Wilson coefficient for the Landau-gauge gluon condensate was needed 
to describe the behaviour of data above $p \simeq 4$ GeV and up to $p \simeq 7$ GeV 
(see next Fig.~{\ref{fig:LambdaMS}). 

\begin{figure}[h]
  \begin{center}
    \includegraphics[width=8cm]{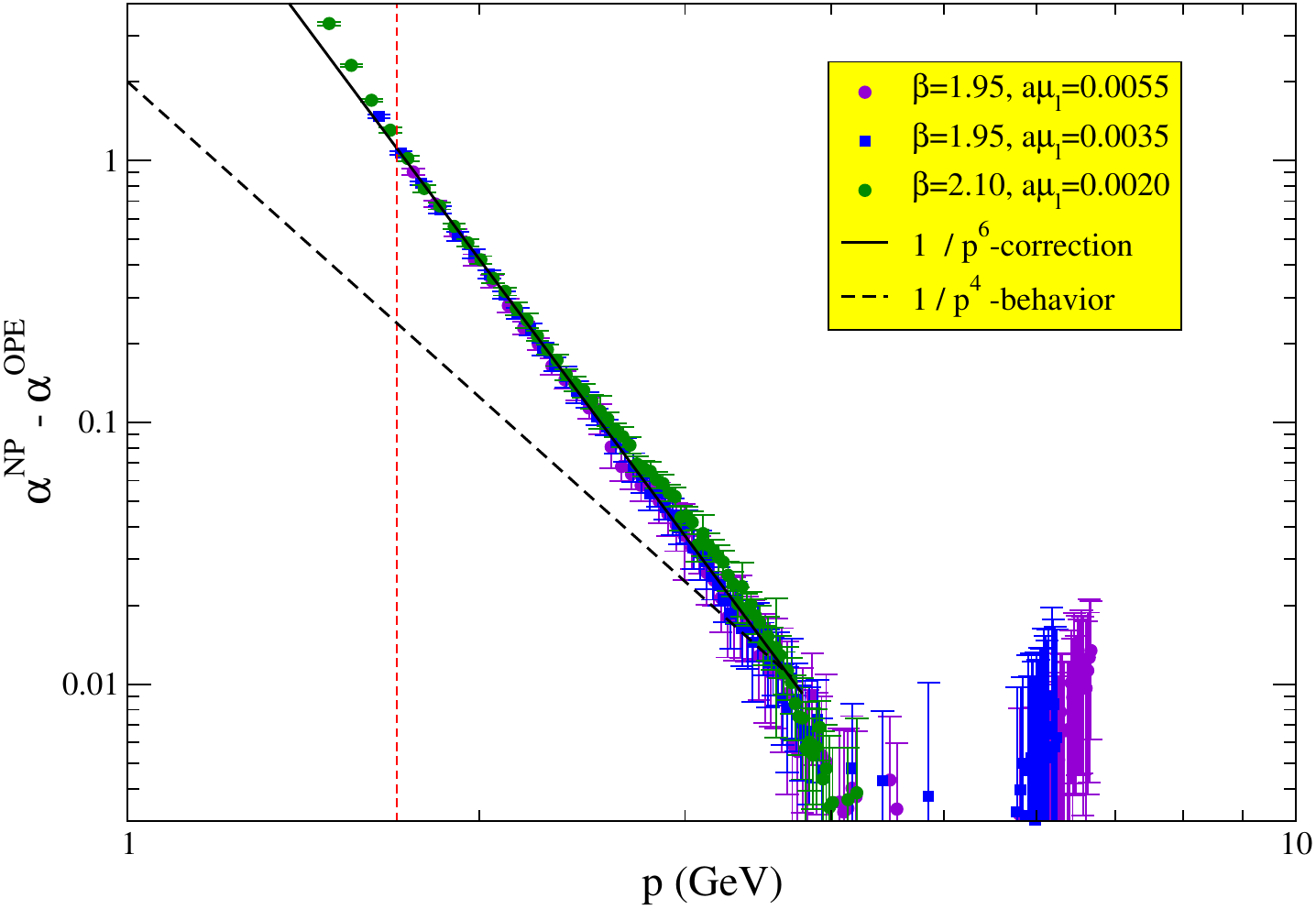}
  \end{center}
\vskip -0.5cm
\caption{\small 
The departure of lattice data from the leading non-perturbative 
OPE prediction for the running coupling plotted in logarithmic scales, in terms of the momentum 
manifestly shows a next-to-leading $1/p^6$ behaviour; the dashed red line stands for the momentum 
scale, $p\simeq 1.7$ GeV, below which the lattice data do not follow the $1/p^6$ behaviour any longer.}
\label{fig:plotalphaTPhi}
\end{figure}

Now, in 
Fig.~\ref{fig:plotalphaTPhi}, the impact of higher-power corrections 
is sketched: the plot shows the departure of the lattice data for the Taylor coupling from the 
prediction given by \eq{alphahNP}, plotted in terms of the momentum, 
with logarithmic scales for both axes. The data seem to indicate that the next-to-leading 
non-perturbative correction is highly dominated by an  $1/p^6$ term. 
This might suggest that the $1/p^4$ OPE contributions are negligible when compared with 
the $1/p^6$ ones or that the product of the leading $1/p^4$ terms and the involved Wilson coefficients 
leave with an effective $1/p^6$ behaviour. Anyhow, this implies that we can effectively describe 
the Taylor coupling lattice data for all momenta above $p \simeq 1.7$ GeV with
\beq\label{eq:alphahdp6}
\alpha_T^{d}(p^2) \ = \ \alpha_T(p^2) \ + \ \frac{d}{p^6} \ ,
\eeq
where $d$ is a free parameter to be fitted which we do not attribute to any particular physical 
meaning. 
Other possible {\it ad-hoc} fitting formulas might be also applied and 
this can be thought to induce a systematice error on the determination of 
$\Lambda_{\msbar}$ in the next section. However, the comparison of perturbative 
and nonperturbative estimates will show this error not to be larger than 
around 20 MeV.

\section{The strong coupling in $\msbar$ scheme }
\label{sec:alphaMS}

To obtain the $\msbar$ $\Lambda_{\rm QCD}$ from 
$\Lambda_T$ is rather immediate, as the scale-independent $\Lambda_{\rm QCD}$-parameters  
in both Taylor and $\overline{\rm MS}$ schemes are related through~\cite{Blossier:2010ky}
\beq\label{ratTMS}
\frac{\Lambda_{\overline{\rm MS}}}{\Lambda_T} \ = 
\exp{\left(\displaystyle - \frac{507-40 N_f}{792 - 48 N_f}\right)} \ = \ 0.560832
\ .
\eeq
Then, one can numerically invert Eqs.~(\ref{alphahNP},\ref{eq:alphahdp6}) and apply 
\eq{ratTMS} to determine $\Lambda_{\msbar}$ from all the lattice estimates of 
the Taylor coupling at any available momenta. 
$\Lambda_{\msbar}$ from different momenta must only differ   
by statistical fluctuations, provided that Eqs.~(\ref{alphahNP},\ref{eq:alphahdp6})
properly describes lattice data at those momenta. Thus, the parameters $g^2\langle A^2 \rangle$ 
and $d$ are to be fixed such that a constant fits with the minimum $\chi^2/$d.o.f. 
to the $\Lambda_{\msbar}$ results obtained by the inversion of Eqs.~(\ref{alphahNP},\ref{eq:alphahdp6}).   
This is the {\it plateau} method applied in Fig.~\ref{fig:LambdaMS}, which is 
equivalent to fit directly Eqs.~(\ref{alphahNP},\ref{eq:alphahdp6}) to the Taylor 
coupling lattice data, as done in Fig.~\ref{fig:plotalphaT}. The best-fit parameters can 
be found in Tab.~\ref{tab:bestfit}. The best {\it plateau} with \eq{eq:alphahdp6} 
is obtained for $\Lambda_{\msbar}=0.324(17)$ GeV over a fitting window ranging 
from $p=1.7$ GeV up to $p=6.8$ GeV, where $\chi^2/\mbox{\rm d.o.f.}=146.9/516$; while 
$\chi^2/\mbox{\rm d.o.f.}=106.7/329$ over $4.1 < p < 6.8$ GeV for $\Lambda_{\msbar}=0.316(13)$ 
with \eq{alphahNP}. 
For the sake of comparison, we also estimate $\Lambda_{\msbar}$ by 
inverting~\eq{alphahNP} with $g^2\langle A^2 \rangle = 0$. A {\it plateau} is then 
possible for a narrow window only including the highest momenta; 
as for $5.5 < p < 6.8$ GeV, where we obtain $\Lambda_{\msbar}=0.351(11)$ GeV 
with $\chi^2/\mbox{\rm d.o.f.}=107.2/154$. 
Indeed, these last estimates clearly show a systematic non-flat behaviour that can be pretty well explained as described in the caption.

\begin{figure}[h]
  \begin{center}
    \includegraphics[width=8cm]{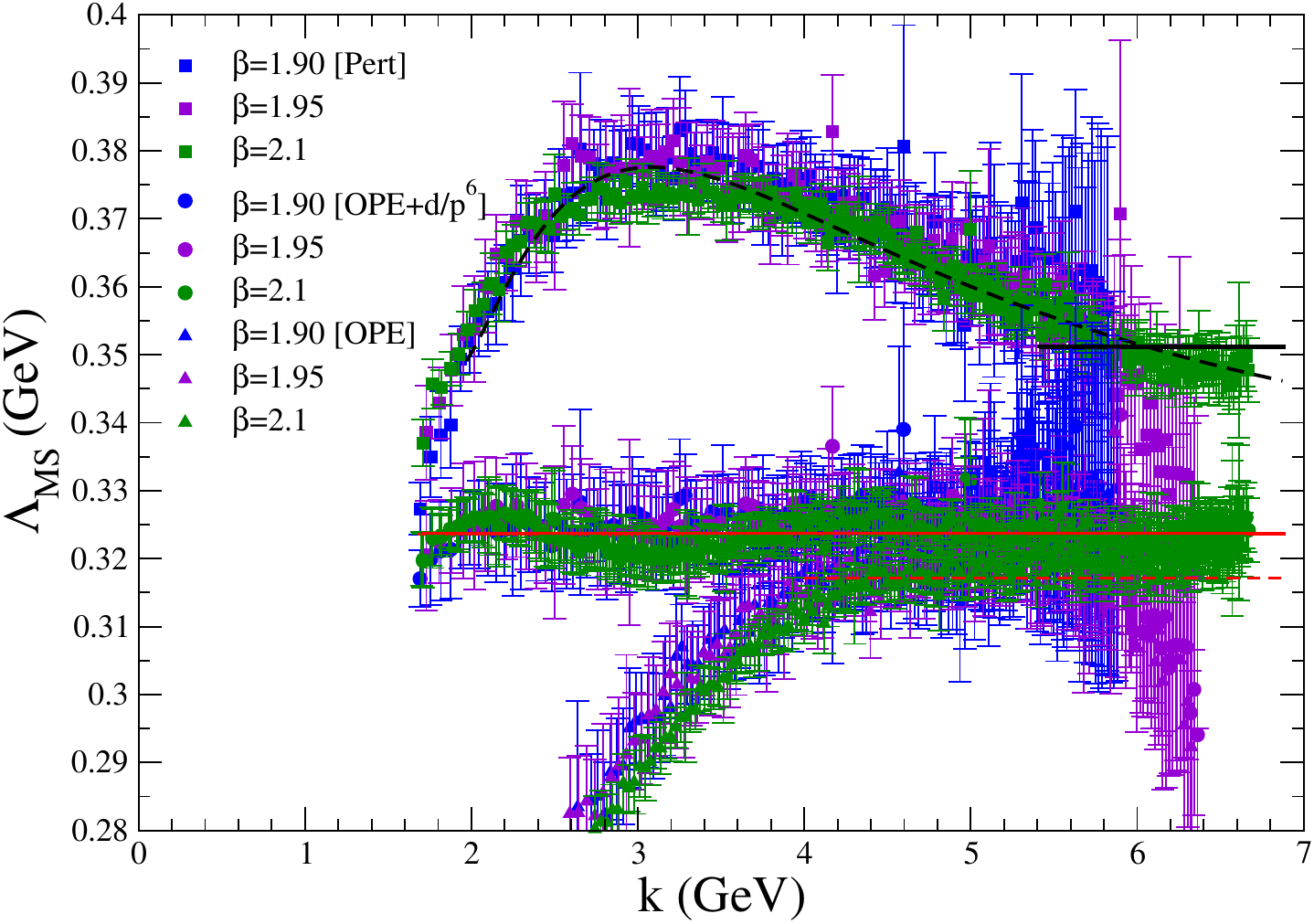} %
  \end{center}
\vskip -0.5cm
\caption{\small $\Lambda_{\msbar}$ obtained by applying the {\it plateau} method 
to the lattice data labelled in the plot. Red solid/dashed line 
corresponds to the {\it plateau} for $\Lambda_{\msbar}$ obtained with 
Eq.~(\ref{eq:alphahdp6})/(\ref{alphahNP}). The black solid is 
for \eq{alphahNP} with $g^2 \langle A^2 \rangle = 0$, while 
black dashed corresponds to 
evaluate first \eq{eq:alphahdp6} with the best-fitted parameters  
in Tab.~\ref{tab:bestfit} and take then the resulting $\alpha_T$ 
to obtain $\Lambda_{\msbar}$ by inverting \eq{alphahNP} with 
$g^2 \langle A^2 \rangle = 0$. 
}
\label{fig:LambdaMS}
\end{figure}
  
\begin{figure}[h]
  \begin{center}
    \includegraphics[width=8cm]{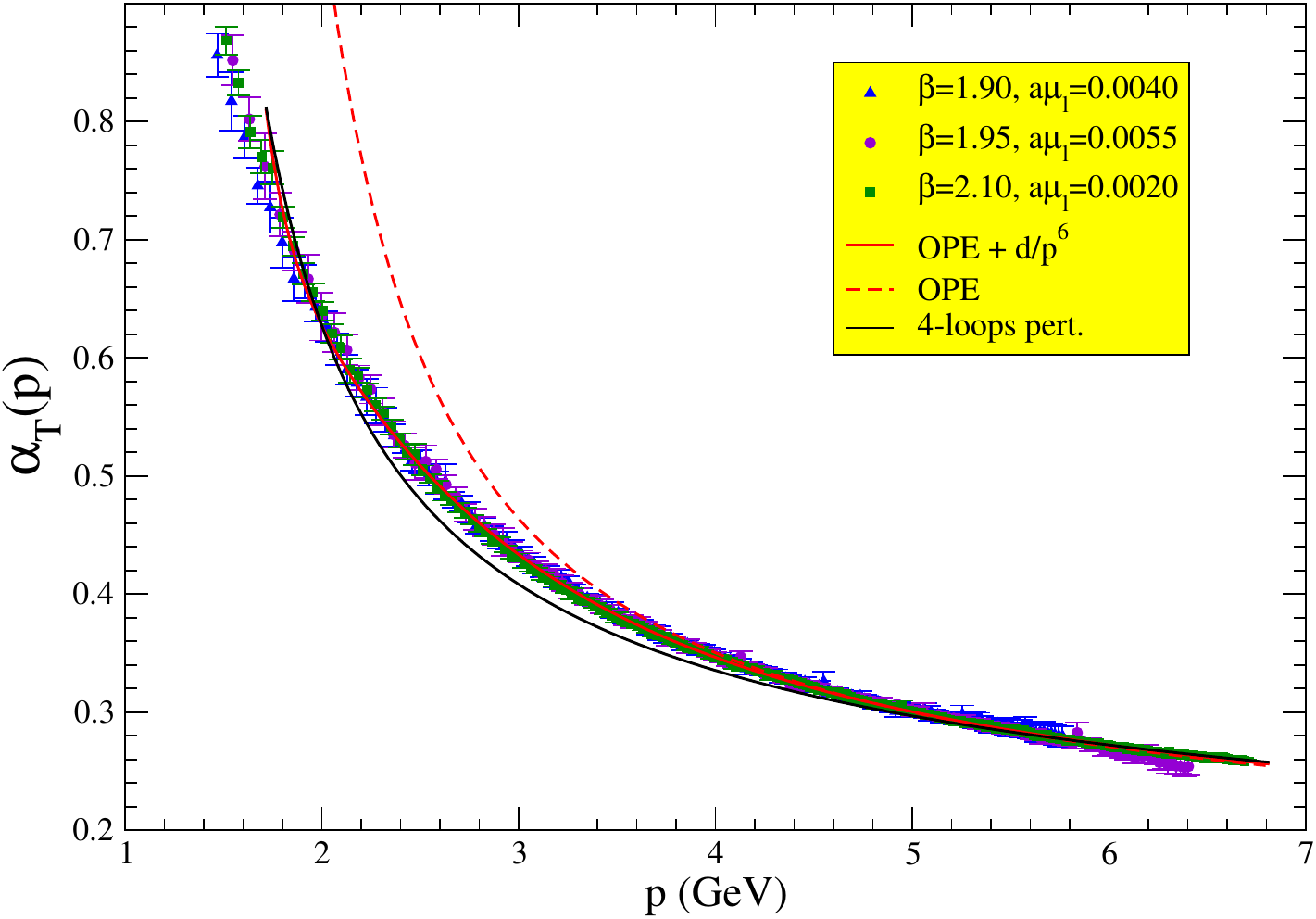} %
  \end{center}
\vskip -0.5cm
\caption{\small \eq{alphahNP} (red dashed) and \eq{eq:alphahdp6} (red solid) for 
the parameters in Tab.~\ref{tab:bestfit} fitted to 
the lattice data for $\alpha_T$ defined by \eqref{alpha}. The black line 
is for \eq{alphahNP} with $g^2 \langle A^2 \rangle=0$.
}
\label{fig:plotalphaT}
\end{figure}

The $\msbar$ running coupling can be obtained again by the integration 
of $\beta$-function, with the coefficients now in $\msbar$-scheme for ${\rm N}_f=4$. 
Thus, we can apply the two estimates of $\Lambda_{\overline{\rm MS}}$, that can be found 
in Tab.~\ref{tab:bestfit}, to run the coupling down to the scale of $\tau$ mass, 
below the bottom quark mass threshold,  
and compare the result with the estimate from $\tau$ decays~\cite{Bethke:2011tr}, 
$\alpha_{\msbar}(m_\tau^2)=0.334(14)$. 
This will produce, with the 1-$\sigma$ error propagation, 
the two following results at the $\tau$-mass scale: 
$\ams(m_\tau^2)=0.337(8)$ and $\ams(m_\tau^2)=0.342(10)$. If we combine 
both estimates and conservatively add the errors in quadrature, we will be 
left with  
\beq
\ams(m_\tau^2) \ = \ 0.339(13) \ ,
\eeq
in very good agreement with the one from $\tau$ decays. This can be graphically 
seen in the plot of Fig.~\ref{fig:plotalphaMS}.

\begin{table}
\begin{center}
\begin{tabular}{|c|c|c|c|}
\hline
& 
$\Lambda_{\msbar}^{N_f=4}$ (MeV) & 
$g^2\langle A^2 \rangle$ (GeV$^2$) & $(-d)^{1/6}$ (GeV) \\
\hline
\eq{alphahNP}~\cite{Blossier:2011tf} & 
316(13) & 4.5(4) & \\
\hline 
\eq{eq:alphahdp6} & 
324(17) & 3.8(1.0) & 1.72(3) \\
\hline
\end{tabular}
\end{center}
\vskip -0.3cm
\caption{\small The parameters for the best-fit of \eq{alphahNP} 
(see ref.~\cite{Blossier:2011tf}) to lattice data (first row) 
and the same with \eq{eq:alphahdp6} (second row). The conversion to $\msbar$ scheme for 
$\Lambda_{\rm QCD}$ is done by applying \eq{ratTMS}. 
The renormalization point for the gluon condensate is fixed at 
$\mu=10$ GeV. We quote statistical errors obtained 
by applying the jackknife method.}
\label{tab:bestfit}
\end{table}


\begin{figure}[h]
  \begin{center}
    \includegraphics[width=8cm]{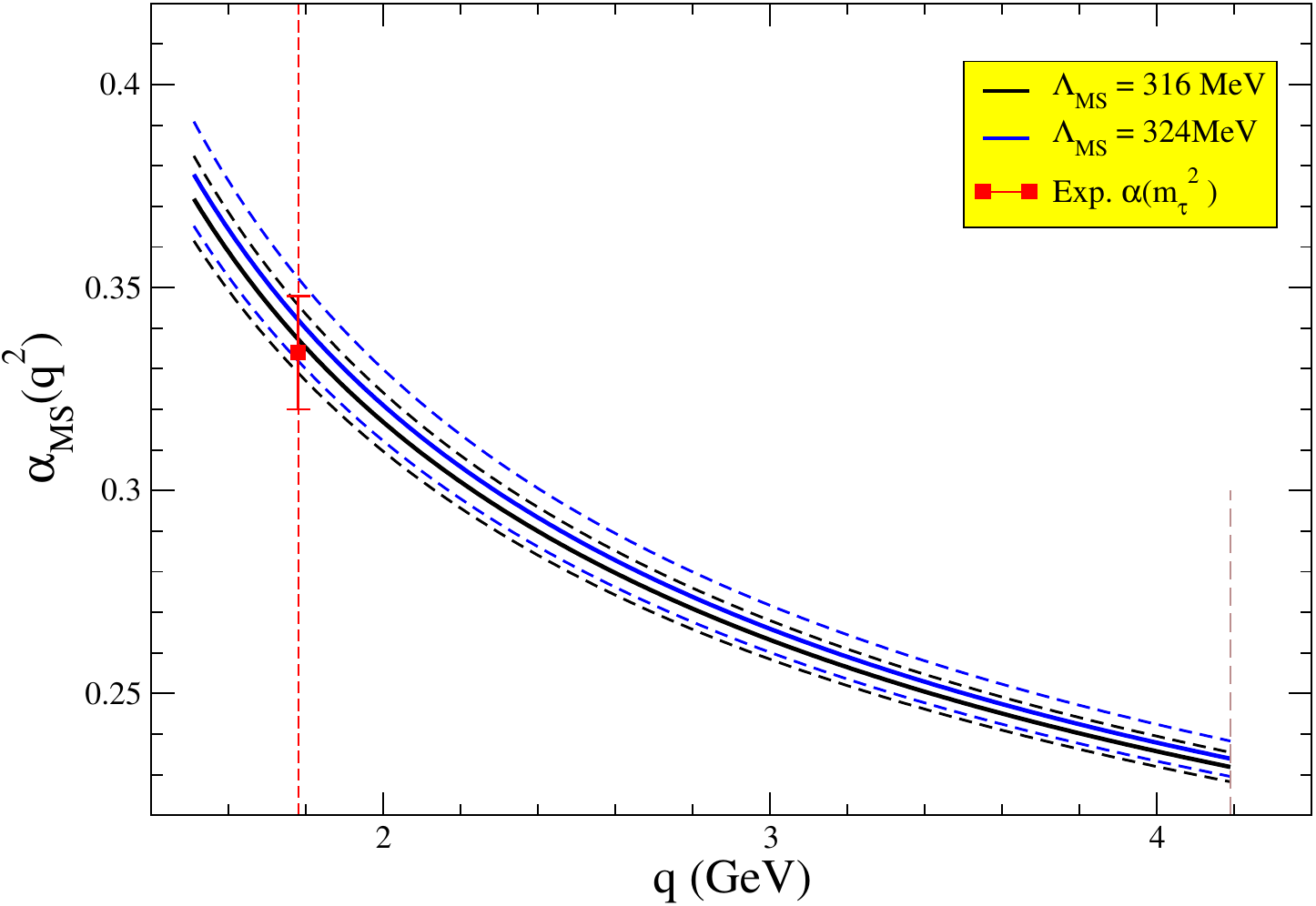}
  \end{center}
\vskip -0.5cm
\caption{\small 
The strong $\msbar$ coupling running for 4 quark flavours
and for $\Lambda_{\overline{\rm MS}}=316$ MeV (black) and 
$\Lambda_{\overline{\rm MS}}=324$ MeV (blue) 
below the bottom mass threshold. The dashed lines represent the one-$\sigma$ statistical 
deviations. The red point stand for the value of $\alpha_{\overline{\rm MS}}(m_\tau^2)$ obtained 
from $\tau$ decays~\cite{Bethke:2011tr}.
}
\label{fig:plotalphaMS}
\end{figure}


The determination of $\ams$ at the $Z^0$ mass scale implies first 
to run up to the MS running mass for the bottom quark, $m_b$, 
with $\beta$-coefficients and $\Lms$ estimated for 4 quark flavours, 
apply next the matching formula~\cite{Nakamura:2010zzi}:
\beq
\ams^{N_f=5}(m_b) = \ams^{N_f=4}(m_b) 
\left( 1 + \sum_n c_{n0} \left(\ \ams^{N_f=4}(m_b)\right)^n \right) ,
\eeq
where the coefficients $c_{n0}$ can be found in ref.~\cite{Chetyrkin:2005ia,Schroder:2005hy}, 
and then run from the bottom mass up to the $Z^0$ mass scale. Thus, 
from our two estimates of $\Lms$, we obtain: $\ams(m_Z^2)=0.1198(9)$ 
and $\ams(m_Z^2)=0.1203(11)$. Again, combining these two results 
and their errors added in quadrature, we will be left with
\beq
\ams(m_Z^2)=0.1200(14) \ ,
\eeq
lying in the same ballpark of lattice results from the PACS-CS collaboration~\cite{Aoki:2009tf}, 
$\ams(m_Z^2)=0.1205(8)(5)$, estimated with 2+1 Wilson improved fermions 
but relatively large pion masses ($\sim 500$ MeV); and from HPQCD~\cite{Davies:2008sw}, 
$\ams(m_Z^2)=0.1183(8)$, with 2+1 staggered fermions. This last is consistently estimated 
from two different methods and 5 different lattice spacings, and is included in the 
2010 {\it world average}~\cite{Nakamura:2010zzi}: $\ams(m_Z^2)=0.1184(7)$ 
(also in the very preliminary 2011 update~\cite{Bethke:2011tr}: $\ams(m_Z^2)=0.1183(10)$). 
Our estimate also agrees well with this {\it world average}, but still 
better with $\ams(m_Z^2)=0.1197(12)$, the average obtained without the lattice HPQCD result and 
without that from DIS non-singlet structure functions~\cite{Blumlein:2006be}, 
$\ams(m_Z^2)=0.1142(23)$, which is more than 2 $\sigma$'s away from most of the other involved estimates. 
However, if the HPQCD lattice result, only including $u,d$, and $s$ quarks is 
replaced by the present one, also including the $c$ quark, 
the {\it world average} would still be consistent: $\ams(m_Z^2)=0.1191(8)$. 

It should be noted that we applied two different fitting 
strategies, taking different fitting windows and studying the impact of higher 
order OPE corrections, and no systematic effect have been observed. Our error analysis 
is based on the jackknife method when we account for the fitted parameters, while the 
statistical uncertainties on the lattice sizes are properly propagated 
into the final estimates. Some other systematic effects (not included in our error budget), 
as those related to the use of the twisted-mass action for the dynamical quarks or to 
the lattice size determination at the chiral limit, could also appear but can be only excluded by the 
comparison with other lattice and experimental estimates.

\section{Conclusions}

We have presented the results for a first computation of the running strong coupling 
from lattice QCD simulations including $u$,$d$,$s$ and $c$ dynamical flavours. We applied 
the procedure of determining the ghost-gluon coupling renormalized in Taylor scheme over 
a large momenta window and then compare this with the perturbative running improved 
via non-perturbative OPE corrections. That procedure has been previously shown 
to work rather well when analysing lattice simulations with ${\rm N}_f$=0 and 2 dynamical 
flavours and so happens here for ${\rm N}_f$=2+1+1. Our estimate for the running strong coupling 
at the $\tau$-mass scale nicely agrees with those from $\tau$-decays and, after being 
properly propagated up to the $Z^0$-mass scale, is pretty consistent with most of the estimates 
applied to obtain the current PDG {\it world average}, although slightly larger than 
the ${\rm N}_f$=2+1 lattice result also used for this average. 


\section*{Acknowledgements} 

We thank the support of Spanish MICINN FPA2011-23781 and 
``Junta de Andalucia'' P07FQM02962 research projects, and  
the IN2P3 (CNRS-Lyon), IDRIS (CNRS-Orsay) and apeNEXT (Rome) 
computing centers. K. Petrov is part of P2IO "Laboratoire of Excellence".


\bibliography{total}

\end{document}